# Defending Grey Attacks by Exploiting Wavelet Analysis in Collaborative Filtering Recommender Systems

Zhihai Yang [1]

[1] Ministry of Education Key Lab for Intelligent Networks and Network Security, Xi'an Jiaotong University, Xi'an, 710049, China
E-mail: zhyang_xjtu@sina.com

*Abstract*—"Shilling" attacks or "profile injection" attacks have always major challenges in collaborative filtering recommender systems (CFRSs). Many efforts have been devoted to improve collaborative filtering techniques which can eliminate the "shilling" attacks. However, most of them focused on detecting push attack or nuke attack which is rated with the highest score or lowest score on the target items. Few pay attention to grey attack when a target item is rated with a lower or higher score than the average score, which shows a more hidden rating behavior than push or nuke attack. In this paper, we present a novel detection method to make recommender systems resistant to such attacks. To characterize grey ratings, we exploit rating deviation of item to discriminate between grey attack profiles and genuine profiles. In addition, we also employ novelty and popularity of item to construct rating series. Since it is difficult to discriminate between the rating series of attacker and genuine users, we incorporate into discrete wavelet transform (DWT) to amplify these differences based on the rating series of rating deviation, novelty and popularity, respectively. Finally, we respectively extract features from rating series of rating deviation-based, novelty-based and popularity-based by using amplitude domain analysis method and combine all clustered results as our detection results. We conduct a list of experiments on both the Book-Crossing and HetRec-2011 datasets in diverse attack models. Experimental results were included to validate the effectiveness of our approach in comparison with the benchmarked methods.

1. INTRODUCTION

Collaborative filtering recommender systems (CFRSs) have become a popular and effective tool for information retrieval especially when users facing information overload. CFRSs also have played an important role in many popular web services such as Netflix, Amazon and etc, which are designed to recommend items based on relevant information for the specific user [3], [5], [11], [14]. However, CFRSs are particularly vulnerable to "shilling" attacks or "profile injection" attacks in which an attacker signs up as a number of "puppet" users and rates fake scores in an attempt to promote or demote the recommendations of specific items by using knowledge of the recommender algorithms [20], [21]. In such attacks, the attackers deliberately insert attack profiles into genuine profiles to change the prediction results which would reduce the trustworthiness of recommendation. The attack profiles indicate the attacker's intention that he wishes a particular item can be rated highest score (called push attack) or lowest score (called nuke attack) [4], [6], [7], [9], [10], [16], [18], [19]. In addition, to avoid being detected easily, the attackers may rate a higher score or lower score on the target items, which generates relatively hidden attack intents in comparison with push attacks or nuke attacks [24], we also call them grey attacks. Of course, they belong to the "shilling" attacks. Therefore, constructing an effective system to defend the attackers and remove them from the CFRSs is crucial.

Although existing work in this area have focused on detecting and preventing the "shilling" attacks or "profile injection" attacks, it has not reached an fully acceptable level of detection performance. In the literature, supervised and semi-supervised methods have focused on the feature extraction of user profiles and train a classifier to perform classification. Burke et al. [3] proposed and studied several attributes derived from user profiles for their utility in attack detection. They employed kNN as their classification approach. However, it was unsuccessful when detecting attacks with small filler size [1] and also suffered from low classifier precision. Then, Williams et al. [15], [24], [28] used several trained classifiers to detect shilling attacks based on extracted features of user profiles. However, they suffered from low accuracy and many genuine profiles are misclassified as attack profiles. Although, [24] used the higher/lower ratings instead of the maximum/minimum ratings to the target item, discussion of detecting such attacks was limited. Moreover, the detection performance was limited when filler size is small. Mobasher et al. [29] employed signatures of attack profiles and were moderately accurate. But, the method suffered from low accuracy in detecting shilling attack. They just focused on individual users and mostly ignored the combined effect of such attackers. In addition, the detection performance was limited when the attack profiles are obfuscated. Wu et al. [30] proposed a semi-supervised hybrid detection method to detect shilling attacks, which combines the naïve Bayesian classifiers and augmented expectation maximization base on several selected metrics. But, their technique suffered from low F-measure [33] when filler size is small. Zhang et al. [31] proposed an ensemble approach to improve the precision of detection by using meta-learning technique. Their proposed method performs better detection performance than the bench marked methods, but, it suffered from low precision when attack size [2] is small. He et al. [32] employed rough set theory to detect shilling attacks though taking features of user profiles as the condition attributes of the decision table. However, their method also suffered from low precision. F. Zhang et al. [17] proposed an online method to detect profile injection attacks based on HHT and SMV. Although their experimental results reported good performance, it was not reached an acceptable precision when the filler size is larger.

---

[1] The ratio between the number of items rated by user u and the number of entire items in the recommender systems.

In addition, unsupervised methods have been proposed for detecting "shilling" attacks, which require certain prior knowledge rather than training samples. To name a few, Bryan et al. [5] proposed a detection method (UnRAP) to identify the attack profiles by introducing the variance adjusted mean square residue. The detection method performs good results on average and random attacks, but it reported poor detection results on bandwagon attack when filler size is small. Lee et al. [25] proposed a hybrid detection method which combines multidimensional scaling approach and clustering-based method to discriminate attackers. The method was effective in detecting average and random and bandwagon attacks, but it was unsuccessful when detecting attacks with small filler size. Mehta et al. [26] exploited similarity structure in attack profiles to separate them from normal genuine profiles by using unsupervised dimensionality reduction. They also proposed two detection methods, PCA-based and PLSA. The algorithm was effective in detecting various attack models, it needs to know the number of attack profiles to be injected in advance. However, their detection performance became worse when the attack profiles are obscured. Zhou et al. [1] proposed a detection technique for identifying group attack profiles, called DeR-TIA, which combines an improved metric based on Degree of Similarity with Top Neighbors (DegSim) and Rating Deviation from Mean Agreement (RDMA). But, it performed poor detection performance when filler size is small. Zhang et al. [19] proposed a spectral clustering method to make recommender systems resistant to the shilling attacks in the case that the attack profiles are highly correlated with each other. Their experimental results reported good performance in random, average and bandwagon attacks. However, it also performed poor precision and recall in AOP attack when attack size is small.

To sum up, we can briefly summarize that it is difficult to improve detection performance for detecting such attacks when filler size or attack size is small. Moreover, few pay attention to the grey attack detection. As an attacker demotes (nuke attack) the target items by rating lowest score or promotes (push attack) the target items by rating highest score, he also can demote or promote the target items by rating lower or higher scores. In fact, the rating behavior of an attacker is very similar to the behavior of a genuine user if the rating of target item is close to the actual rating. For a nuke attack is simply shifting the rating given to the target item from the minimum rating to a rating one step higher, for the push attack, and vice versa [24]. Any profile that includes these ratings is likely to be less suspect. Although a minor change, this has a key effect. Thus, a challenging detection method should not only perform well when attack size or filler size is small, but also effectively defend the grey attacks.

In this paper, we propose an unsupervised attack detection method to make recommender systems resistant to such attacks, which combines discrete wavelet transform (DWT) and EM-based (Expectation-maximization based) clustering method. Since the attackers mimic some rating details of genuine users in shilling attacks, the rating behavior between attackers and genuine users will become more similar, especially for the grey attacks. Although existing features extracted from user profiles can characterize the shilling attacks to some extent, it's difficult to fully discriminate between attack profiles and genuine profiles. Moreover, the above challenges are also significant in grey attacks. Our basic assumption is that we can use DWT to amplify the differences between attack profiles and genuine profiles. In addition, to characterize the features of grey ratings, we use rating deviation of item to address this crucial problem. To construct input series for DWT, we create a list of transformed rating series to address this problem, which exploits the novelty, popularity and rating deviation of item for each user profiles, respectively. Moreover, we employ the empirical model decomposition (EMD) method to extract intrinsic mode functions (IMFs) from the rating series [17]. These can be seen that there are some but not obvious difference between the attack profiles and genuine profiles (as shown in Figures 4-6). To amplify the difference, we further use DWT to transform these series. In essence, a rating series is a non-stationary random series. Therefore, it is very suitable to be processed by DWT which performs well for non-stationary data [17]. After DWT, the differences between attack profiles and genuine profiles become more obvious (as shown in Figures 7-9). Based on the output series of DWT, we extract a list of effective features by using amplitude domain analysis method. And then exploiting EM clustering method to discriminate jointly attackers and genuine users based on the extracted features. In addition, the effectiveness of our proposed approach is validated and benchmark methods are briefly discussed. Experimental results show that our approach performs well for detecting the grey attacks in comparison with the benchmarked methods.

---

[2] The ratio between the number of attackers and genuine users.

The main purposes and major contributions of our paper are summarized as follows:

- To characterize the grey ratings, we use rating deviation of item to discriminate between grey attack profiles and genuine profiles.
- We investigate the grey attacks in collaborative recommender systems and conduct a series of grey attack experiments.
- To extract DWT-based (based on discrete wavelet transform) features, we construct rating deviation-based, popularity-based and novelty-based rating series for each user profile, respectively.
- Based on the outputs of DWT for each rating series, we extract 15 features by using amplitude domain analysis method.
- We conduct a list of experiments on both the Book-Crossing and HetRec-2011 datasets and compare the performance of our proposed method with two benchmarked methods (HHT-SVM and DeR-TIA).

The remaining parts of this paper are organized as follows: Section 2 describes the attack model and introduces the theory of discrete wavelet transform. Our proposed detection method is introduced in Section 3. Experimental results and analysis are presented and discussed in Section 4. Finally, we conclude the paper with a brief summary and directions for future work.

## 2. PRELIMINARIES

In this section, we firstly describe the attack profiles and attack models. In addition, we introduce the theory of discrete wavelet transform to facilitate discussions later.

### 2.1. Attack profiles and attack models

In the literature, "shilling" attacks are classified into two ways: nuke attack and push attack [3]. In order to nuke or push a target item, the attacker should be clearly known the form of an attack profile. The general form of an attack profile is shown in Table 1. The details of the four sets of items are described as follows:

$I_S$: The set of selected items with specified rating by the function $\sigma(i_k^S)$ [13];

$I_F$: A set of filler items, received items with randomly chosen by the function $\rho(i_l^F)$;

$I_N$: A set of items with no ratings;

$I_T$: A set of target items with singleton or multiple items, called single-target attack or multiple-targets attack. The rating is $\gamma(i_j^T)$, generally rated the maximum or minimum value in the entire profiles.

In this paper, we utilize 8 attack models to generate attack profiles. The involved attack profiles and corresponding explanations are listed in Table 2. The details of these attack models in our experiments are described as follows:

1) AOP attack: A simple and effective strategy to obfuscate the Average attack is to choose filler items with equal probability from the top x% of most popular items rather than from the entire collection of items [22].

TABLE I. GENERAL FORM OF ATTACK PROFILES.

| $I_T$ | | | $I_S$ | | | $I_F$ | | | $I_N$ | | |
|---|---|---|---|---|---|---|---|---|---|---|---|
| $i_1^T$ | ... | $i_j^T$ | $i_1^S$ | ... | $i_k^S$ | $i_1^F$ | ... | $i_l^F$ | $i_1^N$ | ... | $i_v^N$ |
| $\gamma(i_1^T)$ | ... | $\gamma(i_j^T)$ | $\sigma(i_1^S)$ | ... | $\sigma(i_k^S)$ | $\rho(i_1^F)$ | ... | $\rho(i_l^F)$ | null | ... | null |

TABLE II. ATTACK SCHEMES.

| Attack Model | $I_S$ | | $I_F$ | | $I_N$ | $I_T$ (push/nuke/grey) |
|---|---|---|---|---|---|---|
| | *Items* | *Rating* | *Items* | *Rating* | | |
| AOP | null | | x-% popular items, ratings set with normal dist around item mean. | | null | $r_{max}/r_{min}/r_{grey}$ |
| Random | null | | randomly chosen | system mean | null | $r_{max}/r_{min}/r_{grey}$ |
| Average | null | | randomly chosen | item mean | null | $r_{max}/r_{min}/r_{grey}$ |
| Bandwago (average) | popular items | $r_{max}/r_{min}$ | randomly chosen | item mean | null | $r_{max}/r_{min}/r_{grey}$ |
| Bandwagon (random) | popular items | $r_{max}/r_{min}$ | randomly chosen | system mean | null | $r_{max}/r_{min}/r_{grey}$ |
| Segment | segmented items | $r_{max}/r_{min}$ | randomly chosen | $r_{min}/r_{max}$ | null | $r_{max}/r_{min}/r_{grey}$ |
| Reverse Bandwagon | unpopular items | $r_{min}/r_{max}$ | randomly chosen | system mean | null | $r_{max}/r_{min}/r_{grey}$ |
| Love/Hate | null | null | randomly chosen | $r_{min}/r_{max}$ | null | $r_{max}/r_{min}/r_{grey}$ |

2) Random attack: $I_S = \emptyset$ and $\rho(i) \sim N(\bar{r}, \bar{\sigma}^2)$ [13];

3) Average attack: $I_S = \emptyset$ and $\rho(i) \sim N(\bar{r}_i, \bar{\sigma}_i^2)$ [13];

4) Bandwagon (average): $I_S$ contains a set of popular items. And then, we use these items as $I_S$, $\sigma(i) = r_{max}/r_{min}/r_{grey}$ (push/nuke/grey) and $\rho(i) \sim N(\bar{r}_i, \bar{\sigma}_i^2)$ [13];

5) Bandwagon (random): $I_S$ contains a set of popular items, $\sigma(i) = r_{max}/r_{min}/r_{grey}$ and $\rho(i) \sim N(\bar{r}, \bar{\sigma}^2)$ (nuke/grey) [13];

6) Segment attack: $I_S$ contains a set of segmented items, $\sigma(i) = r_{max}/r_{min}/r_{grey}$ and $\rho(i) = r_{min}/r_{max}/r_{grey}$ (push/nuke/grey) [8];

7) Reverse Bandwagon attack: $I_S$ contains a set of unpopular items, $\sigma(i) = r_{min}/r_{max}/r_{grey}$ (push/nuke/grey) and $\rho(i) \sim N(\bar{r}, \bar{\sigma}^2)$ [9];

8) Love/Hate attack: $I_S = \emptyset$ and $\rho(i) = r_{max}/r_{grey}$ (nuke/grey) [9].

*2.2. Discrete wavelet transform*

Discrete wavelet transform (DWT) has been recognized as a natural wavelet transform for discrete time signals. Both time and scale parameters are discrete. For a discrete-time sequence x[n], n∈Z, DWT is defined by discrete-time multi-resolution decomposition which could be computed by Mallat pyramidal decomposition algorithm (as shown in Equations (1)-(3)) [23]. However, since half the frequencies of the signal have now been removed, half the samples can be discarded according to Nyquist's rule. The filter outputs are then sub-sampled by 2 (Mallat's and the common notation is the opposite, g- high pass and h- low pass):

$$A_n^0 = x[n], \ n \epsilon N \quad (1)$$

$$A_n^i = \sum_{k \epsilon Z} g(k - 2n) A_k^{i-1}, \ i = 1, 2, \dots, L \quad (2)$$

$$D_n^i = \sum_{k \epsilon Z} h(k - 2n) A_k^{i-1}, \ i = 1, 2, \dots, L \quad (3)$$

where h and g are impulse responses of high-pass filter H and low-pass filter G, respectively. $\{A_n^i\}$ and $\{D_n^i\}$ are scale sequence and wavelet sequence of $2^{-i}$ scale. L is the maximum possible scale of the discrete signal x[n]. The signal is also decomposed simultaneously using a high-pass filter. The outputs give the detail coefficients (from the high-pass filter) and approximation coefficients (from the low-pass) as shown in Figure 1. It is important that the two filters are related to each other and they are known as a quadrature mirror filter.

DWT of a signal is calculated by passing it through a series of filters. The decomposition is repeated to further increase the frequency resolution and the approximation coefficients decomposed with high and low pass filters and then down-sampled (see Figure 2). This is represented as a binary tree with nodes representing a sub-space with different time-frequency localization. And the tree is known as a filter bank.

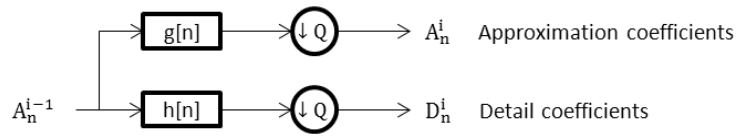

Figure 1. Block diagram of filter analysis.

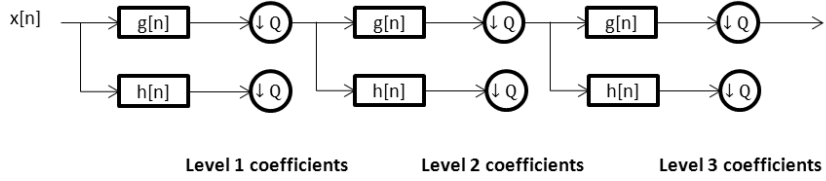

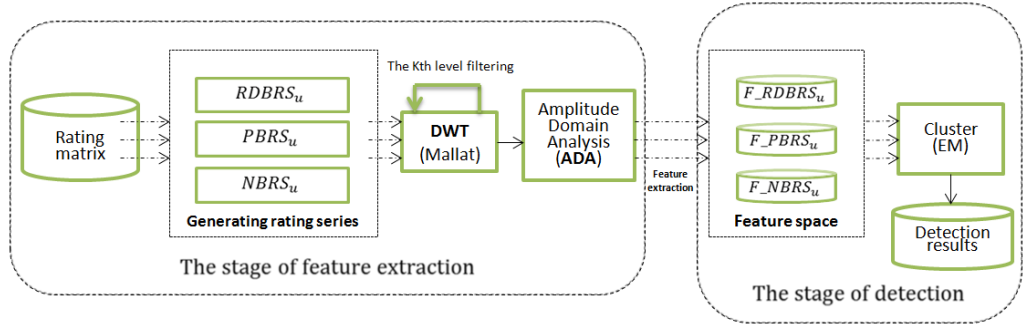

Figure 2. K (k greater than or equal to 1) levels of filter bank.

Figure 3. The framework of our proposed method which consists of two stages: the stage of feature extraction and the stage of detection.

3. OUR PROPOSED APPROACH

In this section, we firstly introduce the framework of our proposed approach. And then we give several definitions of rating series used in this paper. Finally, we briefly describe our detection method.

*3.1. The framework*

As shown in Figure 3, our proposed algorithm consists of two stages: the stage of feature extraction and the stage of detection. At the stage of feature extraction, the feature is extracted one by one from user profiles by using the proposed feature extraction method (see subsection 3.2). Inspired from previous studies (Zhang et al. [17]), we incorporate into two concepts: Empirical Mode Decomposition (EMD) and Intrinsic Mode function (IMF). EMD is an adaptive and highly efficient decomposition method and is also a necessary step to reduce any given data into a collection of intrinsic mode functions (IMF) where the DWT analysis can be applied. As we all know, DWT is a method for analyzing non-stationary data, since the rating series are non-stationary data. The IMF is defined as a function that satisfies the following requirements:

1) In the whole data set

    The number of extreme and zero-crossings must either be equal or differ at most by one;

2) At any point

    The mean value of the envelope defined by the local maxima and the envelope defined by the local minima is zero.

With this method, rating series can be decomposed into a finite signal and regard the signal as the input of discrete wavelet transform [17], [27]. In our proposed approach, we decompose respectively each user profiles into novelty-based, popularity-based and rating deviation-based rating series as the input signals. And then, the input signals are passed through the series of filters (including low-pass and high-pass filter, as shown in Figure 3.) to generate corresponding output signals. In the process of DWT, we perform one level transformation to get the output signals. Then, by using amplitude domain analysis method to extract features from the output signal. At the stage of detection, based on the extracted features, we respectively use EM method to cluster two groups. Finally, combing the three parts of clustering results to return our detection result.

*3.2. Feature extraction*

Previous studies [17] have disclosed that using the novelty and popularity of items to construct rating series for user profiles implies useful information. Inspired from this research, we investigate using rating deviation of items to construct rating series in order to extract features from grey attack profiles. Novelty [3] in recommendation is focusing on recommending the log-tail items (i.e., less popular items) which is

---

[3] The novelty of an item refers to the degree to which it is unusual with respect to the user's normal tastes.

generally considered to be particularly valuable to users. Popularity of items usually reflects the genuine users' tastes or preferences in collaborative recommender system. By sorting the items according to their novelty, popularity and rating deviation, we can create respectively the rating deviation-based, novelty-based and popularity-based rating series for the user profiles. Firstly, two definitions of the rating deviation are described in the following:

**Definition 1** (Rating Deviation of Items, RDoI).

The $\text{RDoI}_i$ (rating deviation of item i) is defined as follows:

$$\text{RDoI}_i = \begin{cases} |r_{ui} - \bar{r}_i|, & r_{ui} \neq \perp, u \in R_g \\ 0, & r_{ui} = \perp \end{cases}, \qquad (4)$$

where $r_{ui}$ denotes the rating of user $u$ on item $i$. $\bar{r}_i$ is the mean rating of item $i$ in the system. $r_{ui} \neq \perp$ denotes item i is rated by user u, $r_{ui} = \perp$ denotes item $i$ is not rated by user $u$. $R_g$ denotes the set of genuine users in dataset.

**Definition 2** (Rating Deviation-based Rating Series, RDBRS).

Let $\text{RDoI}_i$ denotes the rating deviation of item i. Sort all items in set $I$ (a set of the entire items in the recommender system.) according to $\text{RDoI}_i$ in descending order and let $i = 1,2, \dots, |I|$ denotes the order of items after sorting, where $|I|$ denotes total number of items in the recommender system. The $\text{RDBRS}_u(i)$ [4] is defined as follows:

$$\text{RDBRS}_u(i) = \begin{cases} 1, & r_{u,i} \neq \perp \text{ and } (i = 1 \text{ or } \text{RDNRS}_u(i-1) \neq 1), \\ -1, & r_{u,i} = \perp \text{ and } (i = 1 \text{ or } \text{RDNRS}_u(i-1) \neq -1), \\ 0, & \text{otherwise.} \end{cases} \qquad (5)$$

where zero value is used to meet the requirements of extreme for DWT. $r_{u,i} \neq \perp$ denotes item i is rated by user u. $r_{u,i} = \perp$ denotes item $i$ is not rated by user $u$.

**Novelty of Items, NoI**

The $\text{NoI}_i$ (novelty of item i) is defined as follows:

$$\text{NoI}_i = \frac{1}{|R_g|} \sum_{u \in R_g, r_{u,i} \neq \perp} \text{NoI}_{u,i}, \qquad (6)$$

where $\text{NoI}_{u,i}$ denotes the novelty of item $i$ for user u [17].

$$\text{NoI}_{u,i} = \frac{1}{|N_j|} \sum_{u \in R_g, r_{u,j} \neq \perp} (1 - \text{simi}(i,j)) \qquad (7)$$

where $N_j$ denotes the number of users who rate on item j. $R_g$ denotes the set of genuine users in dataset. $\text{simi}(i, j)$ (Jaccard coefficient) denotes the similarity between item $i$ and item $j$, which can be calculated as follows:

$$\text{simi}(i,j) = \frac{|V_i \cap V_j|}{|V_i \cup V_j|} \qquad (8)$$

where $V_i$ is set of users that rated by item i, $V_j$ is the set of users that rated by item j. If both $V_i$ and $V_j$ are empty, we define $\text{simi}(i,j) = 0$. Clearly, $0 \leq \text{simi}(i,j) \leq 1$.

**Novelty-based Rating Series, NBRS**

Let $\text{NoI}_i$ denotes the novelty of item i. Sort all items in set I according to $\text{NoI}_i$ in descending order and let $i = 1,2, \dots, |I|$ denotes the order of items after sorting. The novelty-based rating series of user u, $\text{NBRS}_u(i)$ is defined as follows:

$$\text{NBRS}_u(i) = \begin{cases} 1, & r_{u,i} \neq \perp \text{ and } (i = 1 \text{ or } \text{NBRS}_u(i-1) \neq 1), \\ -1, & r_{u,i} = \perp \text{ and } (i = 1 \text{ or } \text{NBRS}_u(i-1) \neq -1), \\ 0, & \text{otherwise.} \end{cases} \qquad (9)$$

where zero value is used to meet the requirements of extreme for DWT [17].

**Popularity of Items, PoI**

---

[4] The rating deviation-based rating series of user $u$.

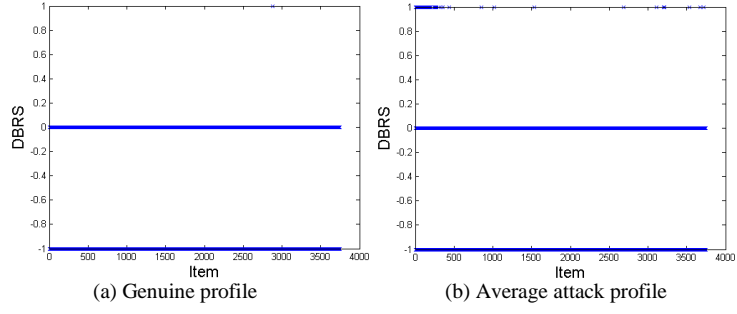

(a) Genuine profile  (b) Average attack profile

Figure 4. Rating Deviation-based rating series. (a) The signal of a genuine profile before DWT; (b) The signal of a average attack profile before DWT.

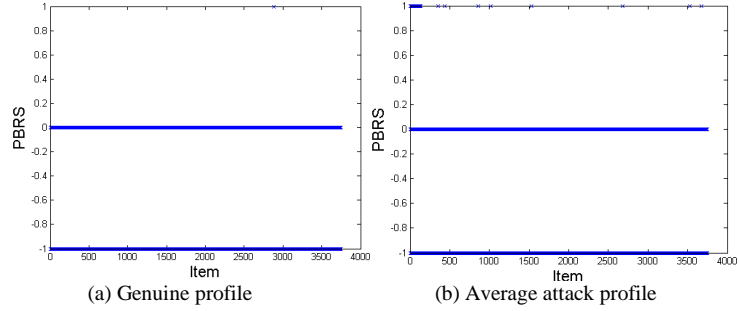

(a) Genuine profile  (b) Average attack profile

Figure 5. Popularity-based rating series. (a) The signal of a genuine profile before DWT; (b) The signal of a average attack profile before DWT.

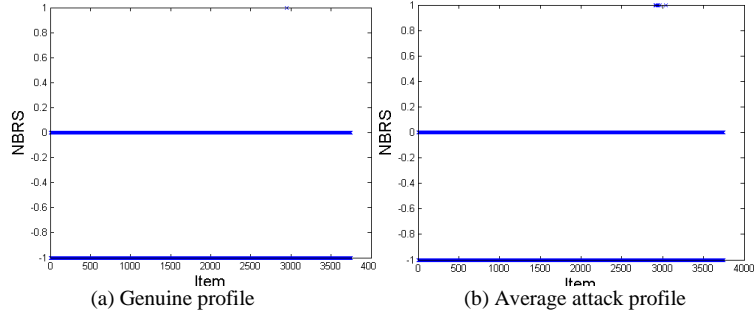

(a) Genuine profile  (b) Average attack profile

Figure 6. Novelty-based rating series. (a) The signal of a genuine profile before DWT; (b) The signal of a average attack profile before DWT.

The popularity of item i, $PoI_i$, is defined as the number of ratings given to item i by genuine users in data set [17].

**Popularity-based Rating Series, PBRS**

Let $PoI_i$ denotes the popularity of item i. Sort all items in set I according to $PoI_i$ in descending order and let $i = 1,2,…,|I|$ denotes the order of items after sorting. The popularity-based rating series of user u, $PBRS_u(i)$, is defined as follows:

$$PBRS_u(i) = \begin{cases} 1, & r_{u,i} \neq \perp \text{ and } (i = 1 \text{ or } PBRS_u(i-1) \neq 1), \\ -1, & r_{u,i} = \perp \text{ and } (i = 1 \text{ or } PBRS_u(i-1) \neq -1), \\ 0, & \text{otherwise.} \end{cases} \quad (10)$$

where zero value is used to meet the requirements of extreme for DWT [17].

To show the difference between genuine and attack profiles in rating series, we give examples of the novelty-based, popularity-based and rating deviation-based rating series in Figures 4-6. These rating series are constructed by the genuine profiles and the average attack profiles (take average attack for example). The genuine profiles are selected from the Book-Crossing dataset. As shown in Figures 4-6, there are very little difference between the genuine and average attack profiles in rating series. We can observe that the RDBRS for the genuine profile barely changed from starting position to ending position in compared to the RDBRS of the average attack profile decreased gradually for the rating deviation-based rating series. For the popularity-based rating series, the PBRS for the genuine profile barely changed with the item increased while the PBRS of the average attack profile decreased gradually. And for the novelty-based rating series, the NBRS for genuine profile also almost remain unchanged with the item increased, while the NBRS of the average attack profile show characteristics of more concentrated. As mentioned above, it is difficult to discriminate between genuine profiles and attack profiles regardless of using Rating Deviation-based, Popularity-based and Novelty-based rating series. To amplify the difference between genuine profiles and attack profiles, we use DWT to transform the rating series in order to extract features from output signal by using amplitude domain analysis method.

After K (k greater than or equal to 1) level discrete wavelet transform (as shown in Figure 2), we can get the local properties, which passes a series low-pass filters to obtain an approximation coefficients. As shown in Figures 7-9, we can observe that there is a more significant difference between genuine profiles and average attack profiles on rating series than before using DWT. In Figure 7, the strength of oscillations of genuine profiles show characteristics of more concentrated with the item increased while the strength of oscillations of average attack profile decreased gradually from starting position to ending position. For the popularity-based rating series, the same observations are also clear in Figure 8. And for the novelty-based rating series, we can observe that there is a little difference between the genuine profiles and average attack profiles, although they show characteristics of more concentrated similarly as illustrated in Figure 9.

Let $F\_RDBRS_u$, $F\_PBRS_u$ and $F\_NBRS_u$ denotes the feature vector of user u on the rating deviation-based, novelty-based and popularity-based after DWT, respectively. The proposed feature extraction algorithm is described in algorithm 1. In algorithm 1, from step 1 to step 3 create the rating deviation-based, novelty-based and popularity-based rating series for user u respectively. Step 4 is the process of DWT. Step 5 extract features from approximation parts of rating deviation, popularity and novelty rating series, termed $A\_RD_k$, $A\_P_k$ and $A\_N_k$ by using amplitude domain analysis method. The last step generates a feature space for the stage of detection.

---

**Algorithm 1:** Feature extraction algorithm for user profiles
**Input:** Rating Matrix;
**Output:** $F\_RDBRS_u$, $F\_PBRS_u$ and $F\_NBRS_u$;
**Step 1:** Create rating series $RDBRS_u(i)$ of $u$ by using rating matrix and Equations (4)-(5);
**Step 2:** Create rating series $NBRS_u(i)$ of $u$ by using rating matrix and Equations (6)-(9);
**Step 3:** Create rating series $PBRS_u(i)$ of $u$ by using rating matrix and Equation (10);
**Step 4:** Generate approximation parts $A$ and detail parts $D$ by exploiting Mallat (discrete wavelet transform) algorithm on the rating series of $RDBRS_u(i)$, $PBRS_u(i)$ and $NBRS_u(i)$ by using Equations (1)-(3), respectively;
**Step 5:** Take the K level approximation parts $A\_RD_k$, $A\_N_k$ and $A\_P_k$ from Step 4's output, respectively. And extract features from the approximation parts by using amplitude domain analysis method on $A\_RD_k$, $A\_N_k$ and $A\_P_k$ respectively;
**Step 6:** Generate and return the feature space $F\_RDBRS_u$, $F\_PBRS_u$ and $F\_NBRS_u$ respectively.

---

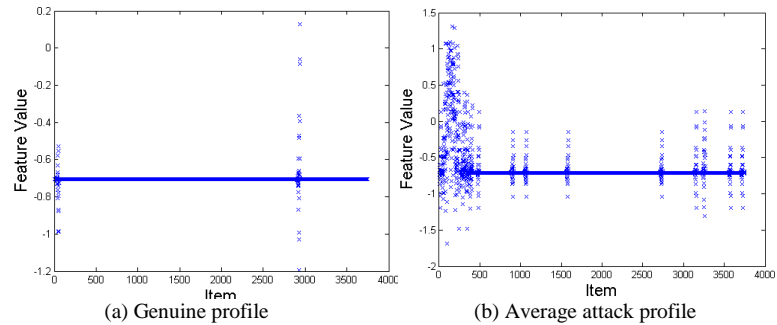

(a) Genuine profile  (b) Average attack profile

Figure 7. The first low-pass output of the rating deviation-based rating series. (a) The signal of a genuine profile after DWT; (b) The signal of a average attack profile after DWT.

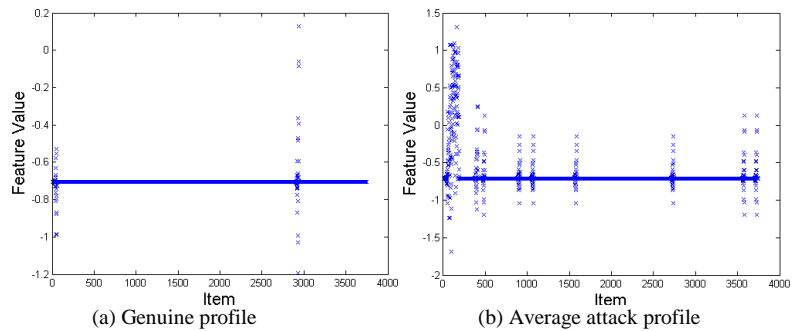

(a) Genuine profile  (b) Average attack profile

Figure 8. The first low-pass output of the popularity-based rating series. (a) The signal of a genuine profile after DWT; (b) The signal of a average attack profile after DWT.

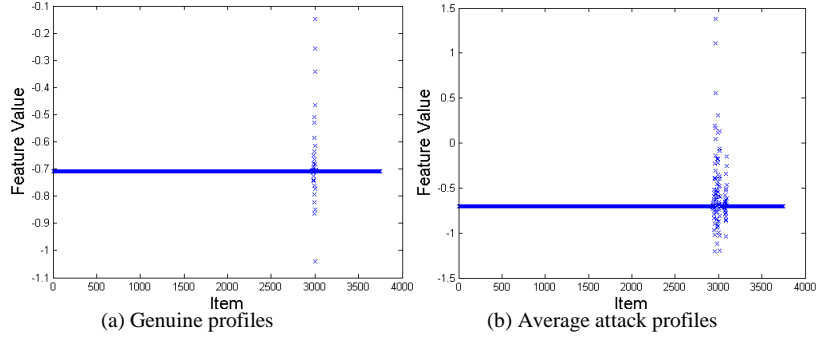

Figure 9. The first low-pass output of the novelty-based rating series. (a) The signal of a genuine profile after DWT; (b) The signal of a average attack profile after DWT.

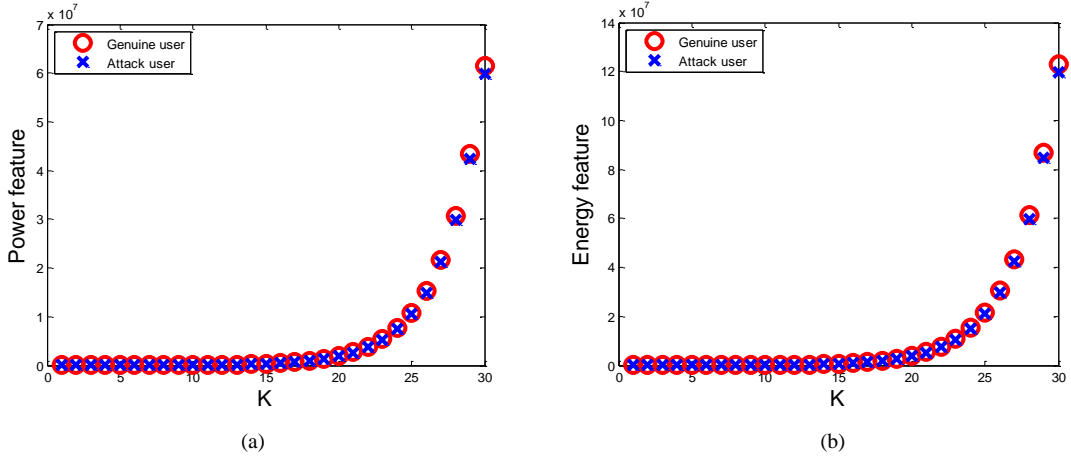

Figure 10. The power feature and the energy feature in different K levels output of discrete wavelet transforms for a genuine user and an attacker. (a) Power features; (b) Energy features.

TABLE III. THE FEATURES OF THE SIGNAL AMPLITUDE DOMAIN AND THEIR DESCRIPTION.

| Features | Equations | Descriptions |
|---|---|---|
| Minimum value | $x_{max} = \max(X)$ | The minimum value of the amplitude of the signal. |
| Maximum value | $x_{min} = \min(X)$ | The maximum value of the amplitude of the signal. |
| Mean value | $\bar{X} = \text{mean}(X)$ | The average value of the amplitude of the signal. |
| Peak value | $x_p = \max(\text{abs}(X))$ | The maximum of the absolute value of the amplitude of the signal. |
| Root mean square value | $X_{rms} = \sqrt{\frac{1}{N}\sum_{i=1}^{N} x_i^2}$ | The root mean square value of the amplitude of the signal. |
| Root mean square amplitude value | $X_r = \left(\frac{1}{N}\sum_{i=1}^{N}\sqrt{|x_i|}\right)^2$ | Represent the energy size of the signal. |
| Absolute mean | $|\bar{X}| = \frac{1}{N}\sum_{i=1}^{N}|x_i|$ | Absolute mean value of the amplitude of the signal. |
| Variance | $\sigma_x^2 = X_{rms}^2 - \bar{X}^2$ | Represent the degree of dispersion of the signal. |
| Skewness | $\alpha = \frac{1}{N}\sum_{i=1}^{N} x_i^3$ | Represent the asymmetry of amplitude probability density function on the vertical axis. |
| Kurtosis | $\beta = \frac{1}{N}\sum_{i=1}^{N} x_i^4$ | Represent the steep degree of the signal curve. |
| Shape factor | $S_f = X_{rms}/|\bar{X}|$ | A shape factor refers to a value that is affected by an object's shape but is independent of its dimensions |
| Crest factor | $C_f = X_{max}/X_{rms}$ | Crest factor is a measure of a waveform, showing the ratio of peak values to the average value. |
| Impulse factor | $I_f = X_{max}/|\bar{X}|$ | Non-dimensional parameter in amplitude domain. |
| Clearance factor | $CL_f = X_{max}/X_r$ | Non-dimensional parameter in amplitude domain. |
| Kurtosis value | $K_v = \beta/X_{rms}^4$ | Non-dimensional parameter in amplitude domain. |

For different types of signal, there are different analysis methods such as time domain analysis, frequency domain analysis and amplitude domain analysis. As shown in Figure 10, we can observe that these are no significant difference between genuine user and attacker with the K (the K level output of DWT) increased, regardless of using the power features or energy features. In this paper, we use amplitude domain analysis to extract features from signals. The details of signal features in amplitude domain are showed in Table 3. We have 15 features to characterize the signal which extracts from the K level (we set K equal to 1 in our work) output of DWT.

*3.3. Detection algorithm*

In order to get better detection performance as far as possible, we combine the rating deviation-based, novelty-based and popularity-based methods to distinguish between genuine profiles and attack profiles. And then, we use EM as the data mining algorithm to detect grey attacks. Let D denotes the set of detection result. The proposed method for detecting grey attacks is described in algorithm 2. In algorithm 2, from step 1 to 3 perform EM algorithm on feature vector $F\_RDBRS_u$, $F\_PBRS_u$ and $F\_NBRS_u$, respectively. Step 4 obtains the set of attackers decided by using the smaller cluster, since the number of attackers less than the number of genuine users in the recommender system. In step 5, we exploit the intersection of the set D_RD, D_P and D_N, and then the detection result D was generated.

---
**Algorithm 2:** Detection algorithm
**Input:** The set of users' feature space $F\_RDBRS_u$, $F\_PBRS_u$ and $F\_NBRS_u$; The number of clusters $k$;
**Output:** The detected result $D$;
**Step 1:** $\{C\_RD_1, C\_RD_2\} \leftarrow \text{EM}(F\_RDBRS_u)$;
**Step 2:** $\{C\_P_1, C\_P_2\} \leftarrow \text{EM}(F\_PBRS_u)$;
**Step 3:** $\{C\_N_1, C\_N_2\} \leftarrow \text{EM}(F\_NBRS_u)$;
**Step 4:** $D\_ARD = \min(C\_RD_1, C\_RD_2)$, $D\_P = \min(C\_P_1, C\_P_2)$, $D\_N = \min(C\_N_1, C\_N_2)$;
**Step 5:** $D \leftarrow \{D | D\_RD \cap D\_P \cap D\_N\}$;
**Return** $D$.

---

## 4. EXPERIMENTS AND ANALYSIS

In this section, we firstly show the experimental data and settings on two real-world datasets. Then, we describe the evaluation metrics such as detection rate and false alarm rate for the comparison with our proposed algorithm and two benchmark methods in different attack models. Secondly, we conduct a list of grey attacks experiments to compare the prediction shift in different grey ratings. Finally, we discuss the experimental results.

*4.1. Experimental data and settings*

In our experiments, we use both the Book-Crossing[5] and HetRec-2011[6] datasets. Book-Crossing contains 278,858 users providing 1,149,780 ratings (explicit/implicit) about 271,379 books and each rater had to rate at least 1 books. All ratings are in the form of integral values between minimum value 1 and maximum value 10. The minimum score means the rater dislikes the book, while the maximum score means the rater enjoyed the book. For HetRec-2011 dataset, it consists of 105,000 integer ratings (from 1-10) of 10197 movies from 2113 users. Similarly, the minimum score means the rater dislikes the movie, while the maximum score means the rater enjoyed the movie. In our experiments, we randomly select 800 genuine profiles from the two datasets as the samples of genuine profiles, respectively.

The attack profiles indicate the attackers intention that he wishes a particular item can be rated the highest or lowest, higher or lower rating. In this paper, we just detect the nuke attacks and their grey attacks, push attacks can be detected in the analogous manner. For each attack model (as shown in Table 2), we respectively generate nuke and grey attack profiles according to the corresponding attack models with diverse attack sizes {3%, 7%, 12%, 17%, 22%, 27%, 32%, 37%, 42%, 47%} and filler sizes {1%, 1.7%, 2.5%, 5%, 6.7%, 8%, 10%}. In addition, to ensure the rationality of the results, the target item is randomly selected for these attack profiles. Especially in Table 2, the $r_{grey}$ is the grey rating on target items rated by lower scores such as 1, 3, 5 and 7. Moreover, for bandwagon (random and average) attacks, we select randomly 10 items from popular items as the selected items which are rated by more than 200 users in the sampled dataset. For segment attack, we select 5 top popular items as the segmented items which are rated with the highest number of users. And for reverse bandwagon attack, we randomly choose 10 items as the selected items which are rated by one user in the sampled dataset.

---
[5] http://www.informatik.uni-freiburg.de/~cziegler/BX/
[6] http://grouplens.org/datasets/hetrec-2011/

To create a training set for the benchmarked method HHT-SVM, we generate attack profiles by using 8 attack models with different filler sizes {1%, 1.7%, 2.5%, 5%, 6.7%, 8%, 10%}, respectively. Besides, to balance the proportion between genuine profiles and attack profiles in the training set, we construct 15 attack profiles for each of the aforementioned attack models corresponding to the filler sizes. Thus, the training dataset consists of 800 genuine profiles and 840 attacker profiles. For test sets, we randomly select 800 genuine profiles from the remaining 278,058 genuine profiles in the Book-Crossing dataset. Furthermore, we generate respectively attack profiles by exploiting 8 attack models with diverse attack sizes {3%, 7%, 12%, 17%, 22%, 27%, 32%, 37%, 42%, 47%} and filler sizes {1%, 1.7%, 2.5%, 5%, 6.7%, 8%, 10%}. And then, the generated attack profiles are respectively inserted into the sampled genuine profiles to construct our test datasets. Therefore, we have 560 (8*10*7) test datasets including 8 attack models, 10 different attack sizes and 7 different filler sizes. For the HetRec-2011 dataset, we generate attack profiles in the same way. Notice that, these process is repeated 10 times and the average value of detection results are reported for the experiments. All numerical studies are implemented using MATLAB R2012a on a personal computer with Intel(R) Core(TM) i7-4790 3.60GHz CPU, 16G memory and Microsoft windows 7 operating system.

To measure detection performance of the proposed methods, we use detection rate and false alarm rate in our experiments.

$$detection\ rate = \frac{|D \cap A|}{|A|} \quad (11)$$

$$false\ alarm\ rate = \frac{|D \cap G|}{|G|} \quad (12)$$

where D is the set of the detected user profiles, A is the set of attacker profiles, and G is the set of genuine user profiles [11].

*4.2. The prediction shift in grey attacks*

To validate the effectiveness of grey attacks in our work, we conduct a list of experiments in average attack (average attack is taken for example) with diverse attack sizes and filler sizes. The target items rated with grey ratings including 3 and 5 score (these two grey ratings are taken for examples). To measure the deviation between the prediction rating and the actual rating, we use Mean Absolute Error (MAE) and Root Mean Squared Error (RMSE) to evaluate the recommendation precision of algorithm.

$$MAE = \frac{\sum_{u,i}|r_{u,i} - \hat{r}_{u,i}|}{N} \quad (13)$$

$$RMSE = \sqrt{\frac{\sum_{u,i}(r_{u,i} - \hat{r}_{u,i})^2}{N}} \quad (14)$$

where $r_{u,i}$ denotes the actual rating user u gave to item i, $\hat{r}_{u,i}$ denotes the rating user u gave to item i as predicted by a method, and N denotes the number of total ratings in the test set [2], [12], [20].

As shown in Figures 11 and 12, one observation is that MAE and RMSE increased gradually with the filler size and attack size increasing when the grey rating on target items is 5 (in Figure 11) or 3 (in Figure 12). These results indicate that these grey attacks are effective to bias the recommendation results in comparison with no attack (both filler size and attack size are zero in the Figures).

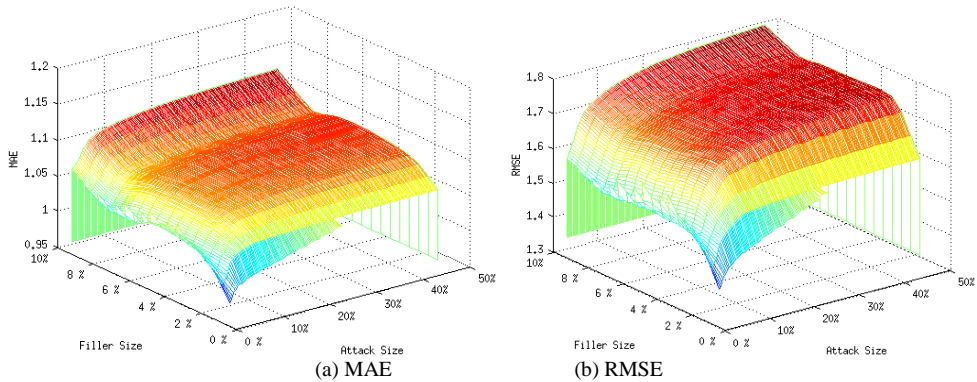

Figure 11. The comparison of MAE and RMSE in different attack sizes and filler sizes. The grey rating is 5. (a) MAE, single-target average attack. (b) RMSE, single-target average attack.

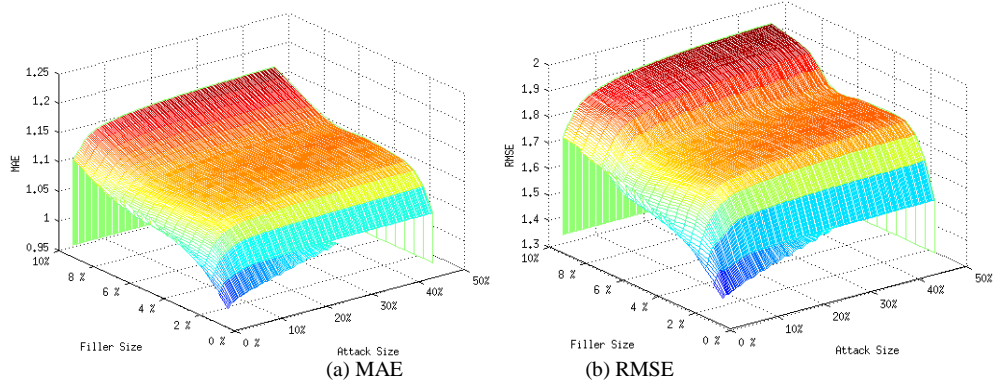

(a) MAE  (b) RMSE

Figure 12. The comparison of MAE and RMSE in different attack sizes and filler sizes. The grey rating is 3. (a) MAE, single-target average attack. (b) RMSE, single-target average attack.

*4.3. Experimental results and analysis*

To validate the detection performance of our proposed method, we employ two benchmarked methods including HHT-SVM [17] and DeR-TIA [1] to demonstrate the outperformance of our method. The details of these two benchmarked methods are described as follows:

- HHT-SVM: An online detection method which combines Hilbert-Huang transforms (HHT) and support vector machine (SVM) and also can operate incrementally. We also use Libsvm 3.18 to generate the classifier. The RBF is used as the kernel function. We set gamma equal to 2 and cost equal to 32 according to the five-cross-validation method.

- DeR-TIA: A technique for detecting attack profiles which uses an improved metric based on degree of similarity with Top Neighbors and rating deviation from mean agreement. We use k-means method in the first phase and set the number of clusters equal to 2. In the second phase, we set the count threshold θ equal to 6.

Take bandwagon (random) attack for example, Figures 13 and 14 demonstrate how each algorithm performs under varying attack sizes and filler sizes, respectively. In the bandwagon (random) attack, a group isolated attackers always provide maximal or minimal or grey rating on a set of items when they are selected as the selected items or the filler items. As shown in Figures 13(a) and 14(a), the detection rate increased gradually and false alarm rate decreased gradually when the attack size increased and the filler size is fixed with 5% (in Figure 13 (a)) and filler size increased and attack size is 17% (in Figure 14 (a)). In addition, we can observe that our method shows significantly better detection performance than HHT-SVM with the attack size increased. This might be attributed to the combination of novelty-based, popularity-based and rating deviation-based rating series adopted by out proposed algorithm. The rating deviation-based strategy calculates a rating offset on a target item which can advantaged identify between the genuine profiles and attack profiles. The second observation is that DeR-TIA shows best performance among the three algorithms. With the attack size increasing, the detection rate almost keeps maximum 100% and the false alarm rate almost keeps minimum 0, except for the early stages (attack size < 17%) as illustrated in Figure 13 (a). The same observations are also clear in Figure 14(a). However, for grey rating, as shown in Figures 13 (b) and 14 (b), we set a grey rating equal to 3 (integer rating from 1-10 in the datasets). Our method shows the best detection performance among the three methods, although the detection rate of our method shows lower than DeR-TIA in the early stage (attack size < 12%) as illustrated in Figure 13 (b). To compare with our proposed method and HHT-SVM, DeR-TIA shows higher false alarm rate than the others. Moreover, the detection rate of DeR-TIA almost remained unchanged with the attack size increased, and similar results can be observed in Figure 14 (b). The results might be attributed to grey rating. The first phase of DeR-TIA can filter out a part of genuine users by using similarity threshold, but it is difficult to capture the suspected profiles which rate grey ratings in their second phase. They defend and remove the suspected users almost depend on the similarity threshold, so they perform lower detection performance. For our proposed method, we pay more attention to the details of the all ratings that rated by a user and explore the top-N items which has sorted by the rating deviation of item in order to characterize the grey ratings.

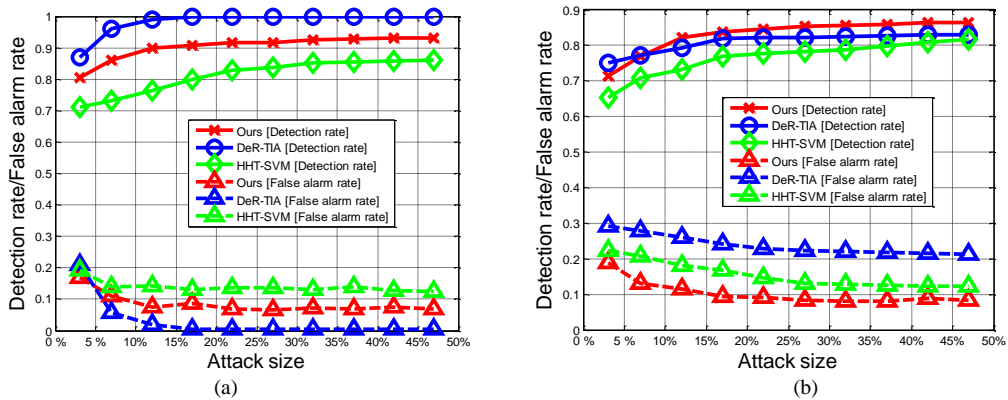

Figure 13. The comparison of detection rate and false alarm rate in different attack sizes. (a) Grey rating is 1, filler size is 5%, single-target bandwagon (random) attack; (b) Grey rating is 3, filler size is 5%, single-target bandwagon (random) attack.

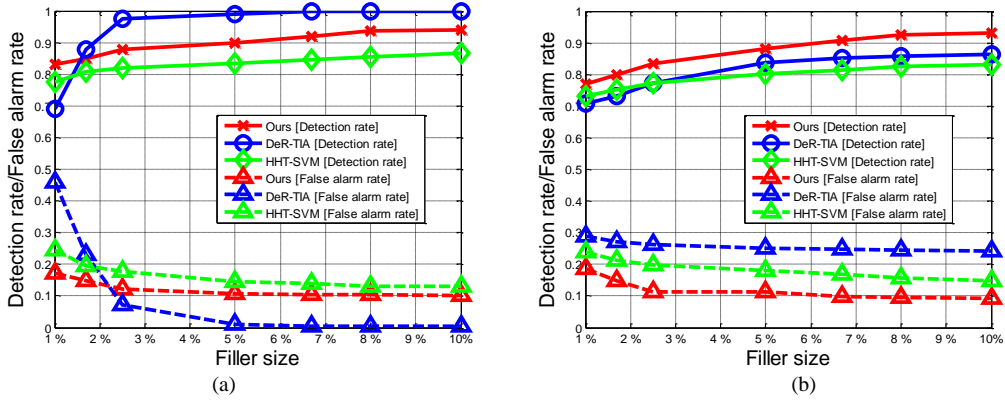

Figure 14. The comparison of detection rate and false alarm rate in different filler sizes. (a) Grey rating is 1, attack size is 17%, single-target bandwagon (random) attack; (b) Grey rating is 3, attack size is 17%, single-target bandwagon (random) attack.

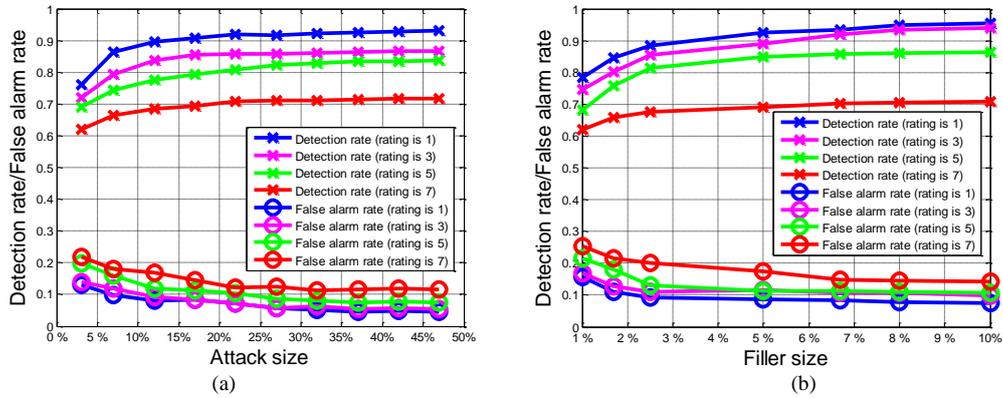

Figure 15. The comparison of detection rate and false alarm rate with different grey ratings in single-target attack. (a) Filler size is 5%, attack size varies in bandwagon (average) attack. (b) Attack size is 17%, filler size varies in bandwagon (average) attack.

To examine the detection performance of our method in bandwagon (random) attack with different grey ratings (take bandwagon (random) attack for example), we conduct a list of experiments with diverse attack sizes and filler sizes. As shown in Figure 15, we perform 4 different grey ratings including 1, 3, 5 and 7 on the target items. One observation is that the detection rate gradually increased and false alarm rate gradually decreased with the attack size increasing (in Figure 15 (a)) or filler size increasing (in Figure 15 (b)). The other observation is that the detection performance gradually performs poor when the grey rating increased from 1 to 7, regardless of different attack sizes and filler sizes. The results may indicate that the grey ratings are close to average rating in the entire system with the grey rating on the target items increasing. The attackers rate an mean rating may show a rating behavior like genuine users, which is difficult to discriminate between attackers and genuine users and shows higher false alarm rate.

TABLE IV. COMPARISON OF THE DETECTION PERFORMANCE OF OUR METHOD WITH TWO BENCHMARKED METHODS.

| Attack models | Methods | Rating | | | | | | | |
|---|---|---|---|---|---|---|---|---|---|
| | | 1 | | 3 | | 5 | | 7 | |
| | | DR | FAR | DR | FAR | DR | FAR | DR | FAR |
| AOP | HHT-SVM | 0.845 | 0.095 | 0.819 | 0.15 | 0.79 | 0.177 | 0.673 | 0.21 |
| | DeR-TIA | 1.0 | 0.005 | 0.715 | 0.185 | 0.734 | 0.225 | 0.707 | 0.275 |
| | Ours | 0.911 | 0.0785 | **0.835** | **0.093** | **0.813** | **0.102** | **0.702** | **0.11** |
| Random | HHT-SVM | 0.819 | 0.12 | 0.765 | 0.15 | 0.7345 | 0.14 | 0.68 | 0.21 |
| | DeR-TIA | 1.0 | 0.0025 | 0.735 | 0.175 | 0.727 | 0.195 | 0.731 | 0.265 |
| | Ours | 0.904 | 0.081 | **0.834** | **0.086** | **0.801** | **0.093** | **0.707** | **0.11** |
| Average | HHT-SVM | 0.873 | 0.1091 | 0.782 | 0.13 | 0.759 | 0.158 | 0.665 | 0.182 |
| | DeR-TIA | 1.0 | 0.0025 | 0.763 | 0.165 | 0.750 | 0.205 | 0.752 | 0.195 |
| | Ours | 0.907 | 0.085 | **0.837** | **0.090** | **0.805** | **0.079** | **0.703** | **0.125** |
| Bandwagon (average) | HHT-SVM | 0.906 | 0.09 | 0.8279 | 0.14 | 0.7869 | 0.16 | 0.675 | 0.19 |
| | DeR-TIA | 1.0 | 0.005 | 0.755 | 0.18 | 0.734 | 0.25 | 0.752 | 0.285 |
| | Ours | 0.935 | 0.0615 | **0.852** | **0.0713** | **0.823** | **0.0682** | **0.705** | **0.115** |
| Bandwagon (random) | HHT-SVM | 0.910 | 0.095 | 0.8179 | 0.13 | 0.8069 | 0.18 | 0.67 | 0.21 |
| | DeR-TIA | 1.0 | 0.005 | 0.747 | 0.165 | 0.735 | 0.205 | 0.750 | 0.27 |
| | Ours | 0.934 | 0.055 | **0.868** | **0.075** | **0.83** | **0.069** | **0.718** | **0.115** |
| Segment | HHT-SVM | 0.897 | 0.0891 | 0.819 | 0.13 | 0.7869 | 0.167 | 0.667 | 0.193 |
| | DeR-TIA | 1.0 | 0.0055 | 0.752 | 0.15 | 0.730 | 0.185 | 0.731 | 0.25 |
| | Ours | 0.915 | 0.075 | **0.846** | **0.08** | **0.815** | **0.086** | **0.70** | **0.11** |
| Reveres bandwagon | HHT-SVM | 0.895 | 0.087 | 0.8179 | 0.125 | 0.796 | 0.145 | 0.66 | 0.195 |
| | DeR-TIA | 1.0 | 0.005 | 0.739 | 0.175 | 0.754 | 0.185 | 0.727 | 0.26 |
| | Ours | 0.933 | 0.065 | **0.868** | **0.075** | **0.815** | **0.0775** | **0.705** | **0.125** |
| Love/Hate | HHT-SVM | 0.849 | 0.105 | 0.807 | 0.135 | 0.7569 | 0.175 | 0.67 | 0.205 |
| | DeR-TIA | 1.0 | 0.0025 | 0.752 | 0.16 | 0.727 | 0.195 | 0.750 | 0.24 |
| | Ours | 0.917 | 0.075 | **0.845** | **0.065** | **0.81** | **0.0785** | **0.717** | **0.135** |

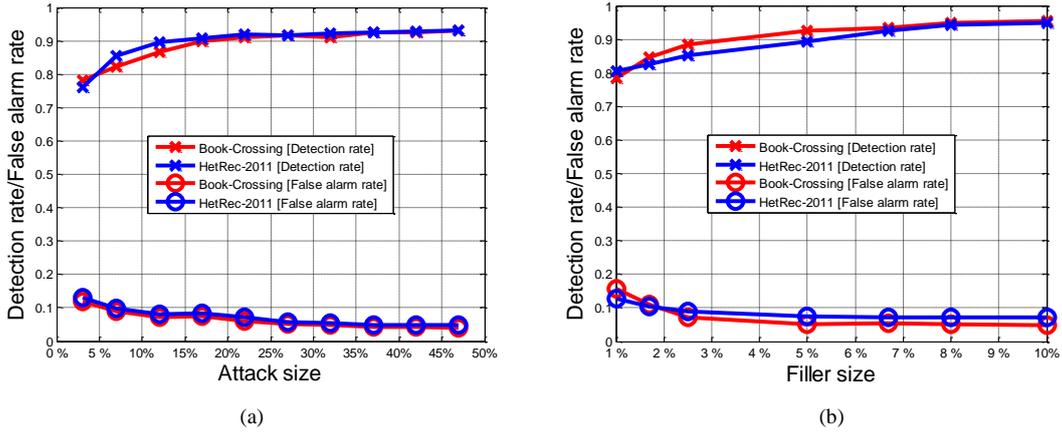

(a)      (b)

Figure 16. Detection rate and false alarm rate in different datasets under single-target segment attack. Grey rating is 3. (a) Filler size is 5% and attack size varies. (b) Attack size is 17% and filler size varies.

To further illustrate the detection performance of our proposed method under different attack models with different grey ratings, we conduct a list of experiments in 8 attack models (such as AOP, random etc) for comparing the performance of our proposed method with HHT-SVM and DeR-TIA. We use 4 different ratings including 1, 3, 5 and 7 score) when filler size is 5% and attack size is 17%. As shown in Table 4, we can observe that the detection rate (DR) of our method reports higher than other two benchmarked methods when the grey rating increasing, except for the grey rating is 1. Similarly, the false alarm rate (FAR) of our method reports lower than others. In addition, the second observation is that the proposed method reports better detection performance under bandwagon (both random and average) and reverse bandwagon attacks in comparison with the other attack models, especially for grey ratings (such as 3, 5 and 7 score). These results may indicate that we combine the rating deviation-based, novelty-based and popularity-based rating series in our method is useful to discriminate difference between grey attack profiles and genuine profiles. The rating deviation-based rating series may easily characterize the grey attacks in comparison with the other two methods.

To evaluate the detection performance of our method in different datasets, we conduct a list of experiments on both the Book-Crossing and HetRec-2011 datasets with different attack sizes (in Figure 16 (a)) and filler sizes (in Figure 16 (b)). Take the segment attack for example, we generate a series of grey attacks when the target items rated with grey rating 3. Figure 16 shows the detection performance of our proposed method in the two different datasets. We can observe that the detection rate increased gradually and the false alarm rate decreased gradually with the attack size or filler size increasing in both the two datasets. Of course, our method is completely data-driven. The experimental results show good detection performance is not invariably mean effective in every datasets.

## 5. CONCLUSIONS AND FUTURE WORK

In this paper, we highlighted the challenges faced by the grey attacks, and then we develop an unsupervised detection approach based on discrete wavelet transform by combing the rating deviation-based, novelty-based and popularity-based rating series. Extensive experiments on the Book-Crossing dataset have demonstrated the effectiveness of the proposed approach. To compare with the benchmarked methods (HHT-SVM and DeR-TIA), our proposed method performs the best detection performance especially for detecting grey attacks. In addition, our method shows higher detection performance than HHT-SVM in the bandwagon (both random and average) and reverse bandwagon attacks. We also conduct a list of experiments on HetRec-2011 dataset to validate the detection performance of our method. Results show that our proposed method also is effective for detecting grey attacks. One of the limitations of our proposed method directly comes from the time consumption, which constructs the signals of rating series. However, it is important for our method to learn new types of attacks incrementally, since they are generated over time in the context of real collaborative recommender systems. In our future work, we intend to extend and improve grey attack detection in the following directions: 1) Considering more attack models such as Power users attack or Power items attack, etc.; 2) We will explore specific and simple method to detect grey attacks and develop better approach to construct the rating series. 3) Extracting more simpler and effective features to characterize grey attack profiles is still an open issue.

## ACKNOWLEDGEMENTS

The research presented in this paper is supported in part by the National Natural Science Foundation (61221063, U1301254), 863 High Tech Development Plan (2012AA011003) and 111 International Collaboration Program, of China.